\documentclass[pra,jmp,amsmath,amssymb,12pt,
reprint,showkeys,showpacs]{revtex4-1}
\usepackage{graphicx}
\usepackage{dcolumn}
\usepackage{bm}
\usepackage[mathlines]{lineno}
\usepackage{amsthm}
\usepackage[bookmarks]{hyperref}
\usepackage{pstricks}
\usepackage{dsfont}
\usepackage{mathrsfs}

\begin{document}

\title[Comments on ``Observation of Fast Evolution in Parity-Time-Symmetric System'']{Comments on ``Observation of Fast Evolution in Parity-Time-Symmetric System''}

\author{F. Masillo}
\email{masillo@le.infn.it}
\affiliation{Dipartimento di Fisica,  Universit\`{a} del Salento}
\affiliation{INFN, Sezione di Lecce, I-73100 Lecce, Italy}
\date{\today}

\begin{abstract}
In the paper ``Observation of Fast Evolution in Parity-Time-Symmetric System'' the authors propose a physical apparatus for the realization of a faster than Hermitian evolution. This last appears  in contrast with the conclusions obtained in our paper ``Some Remarks on  Quantum Brachistochrone''.
We will clarify this apparent contradiction and some problematic aspects of the treatment in \cite{ZHL11}.
\end{abstract}

\pacs{
03.65.-w,
03.65.Yz,
03.65.Ca,
03.67.-a,
03.67.Lx.
}
\keywords{}

\maketitle

The Quantum Brachistochrone Problem can be formulated as follows:

\emph{Given two quantum states $|\psi_i\rangle$ and $|\psi_f\rangle$, we want to find the (time-independent) Hamiltonian $H$ that performs the transformation
    \begin{equation}
    |\psi_i\rangle\rightarrow|\psi_f\rangle=e^{-\frac{i}{\hbar}H\tau}|\psi_i\rangle
    \end{equation}
in the minimal time $\tau$, for a fixed value of the  difference of eigenvalues of $H$.}

This problem was solved in \cite{CHKO06} where it was shown that, if we  put $\omega=|E_+- E_-|$, the minimal time to perform the required transformation is
    \begin{equation}\label{eq.tau}
    \tau=\frac{2\hbar}{\omega}\arccos|\langle\psi_i|\psi_f\rangle|.
    \end{equation}
The same  problem can be formulated allowing the use of pseudo Hermitian Hamiltonians \cite{BBJM06} and its solution gives the apparently paradoxical possibility of operating  a computational process in  an arbitrarily  small time and with limited energy costs \cite{M07}.

 In section $\mathbf{III}$ of \cite{Mas11} we showed that the analysis of the so called transition problem, introduced in \cite{M07d}, induces the impossibility to realize a pseudo Hermitian dynamics of an Hermitian quantum system. This last observation implies that only simulations of pseudo Hermitian (PH)  dynamics by means of open quantum dynamics are possible (note that in the finite dimensional case PT symmetric is equivalent to pseudo Hermiticity (see \cite{SS03} and \cite{M02a})).

\begin{figure}
  % Requires \usepackage{graphicx}
  \includegraphics[width=8cm]{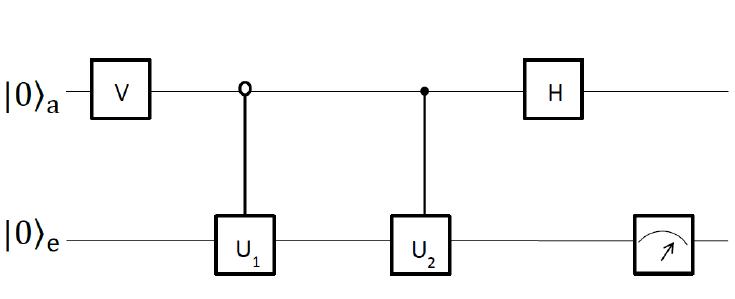}\\
  \caption{The experimental apparatus used in \cite{ZHL11}.}\label{fig1}
\end{figure}

\begin{figure}
  % Requires \usepackage{graphicx}
  \includegraphics[width=8cm]{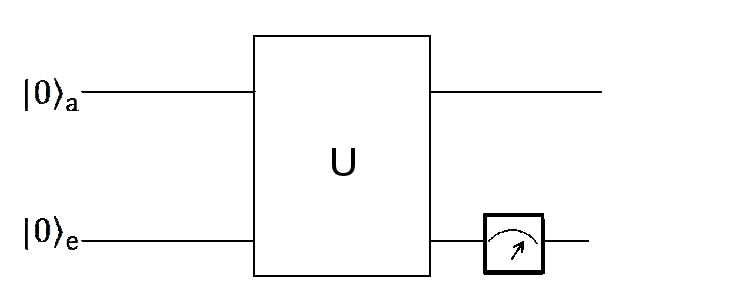}\\
  \caption{A more general apparatus.}\label{fig2}
\end{figure}

In \cite{ZHL11} the authors propose a physical apparatus, whose schematic representation is given in figure  \ref{fig1}, that simulates a PT symmetric dynamics
(see \cite{ZHL11} for the explicit expression of the unitary operators $V,U_1$ and $U_2$).
For our remarks, we need only to note that the apparatus realizes a unitary transformation between the initial and final states (see figure \ref{fig2}).

This apparatus apparently contradicts our analysis in section $\mathbf{IV}$ of \cite{Mas11}, where we showed that a simulation of a pseudo Hermitian dynamics by an Hermitian open dynamics inevitably introduces dissipative effects. In fact  the subdynamics on the work qubit $e$, obtained tracing over the auxiliary space $a$, in the apparatus represented in figure \ref{fig1}, is a trace preserving completely positive map. But using Proposition $\mathbf{IV.1}$ of \cite{Mas11}, this last observation has as immediate consequence that the subdynamics is   a pseudo Hermitian evolution only on a restricted subset of states. Indeed this simulation is only partial and this  contradiction is solved.

More explicitly, we note that  the subdynamics proposed  in \cite{ZHL11} is  not deterministic and so  a non zero  probability that the desired evolution is not realized exists.
In particular  we observe that
\begin{equation}
   U|0_e\rangle\otimes|0_a\rangle=\alpha\frac{e^{-\frac{i}{\hbar}Ht}}{\gamma}|0_e\rangle\otimes|0_a\rangle+\beta|\psi_e\rangle\otimes|1_a\rangle
\end{equation}
where $\gamma=\|e^{-\frac{i}{\hbar}Ht}|0_e\rangle\|$ and $|\beta|^2$ is the probability to obtain the wrong transformation.

Moreover, equation (\ref{eq.tau}) can be used to establish the connection between the physical quantities involved in the transformation:
 \begin{equation}\label{eq.tau2}
    T\geq\frac{2\hbar}{\omega}\arccos|\alpha(\langle 0_e|e^{-\frac{i}{\hbar}HT}|0_e\rangle)|.
    \end{equation}
where $T$ is the time needed to operate the transformation.
If after a time $T$ we obtain that
\begin{equation*}
    \frac{e^{-\frac{i}{\hbar}HT}}{\gamma}|0_e\rangle=|1_e\rangle
\end{equation*}
we have that the faster evolution is due to an increment of $\omega$, and not to the pseudo Hermitian nature of $H$. Moreover because $|\alpha|$ is necessarily  $<1$ we obtain that this simulation  is not efficient with respect to a purely closed Hermitian dynamics.
This means that we can realize a pseudo Hermitian evolution, but, according  to the analysis in sections $\mathbf{IV}-\mathbf{V}$ of \cite{Mas11}, this simulation  has an energy and/or efficiency cost.
 \newpage

\bibliography{comments}

%merlin.mbs apsrev4-1.bst 2010-07-25 4.21a (PWD, AO, DPC) hacked
%Control: key (0)
%Control: author (8) initials jnrlst
%Control: editor formatted (1) identically to author
%Control: production of article title (-1) disabled
%Control: page (0) single
%Control: year (1) truncated
%Control: production of eprint (0) enabled
\begin{thebibliography}{8}%
\makeatletter
\providecommand \@ifxundefined [1]{%
 \@ifx{#1\undefined}
}%
\providecommand \@ifnum [1]{%
 \ifnum #1\expandafter \@firstoftwo
 \else \expandafter \@secondoftwo
 \fi
}%
\providecommand \@ifx [1]{%
 \ifx #1\expandafter \@firstoftwo
 \else \expandafter \@secondoftwo
 \fi
}%
\providecommand \natexlab [1]{#1}%
\providecommand \enquote  [1]{``#1''}%
\providecommand \bibnamefont  [1]{#1}%
\providecommand \bibfnamefont [1]{#1}%
\providecommand \citenamefont [1]{#1}%
\providecommand \href@noop [0]{\@secondoftwo}%
\providecommand \href [0]{\begingroup \@sanitize@url \@href}%
\providecommand \@href[1]{\@@startlink{#1}\@@href}%
\providecommand \@@href[1]{\endgroup#1\@@endlink}%
\providecommand \@sanitize@url [0]{\catcode `\\12\catcode `\$12\catcode
  `\&12\catcode `\#12\catcode `\^12\catcode `\_12\catcode `\%12\relax}%
\providecommand \@@startlink[1]{}%
\providecommand \@@endlink[0]{}%
\providecommand \url  [0]{\begingroup\@sanitize@url \@url }%
\providecommand \@url [1]{\endgroup\@href {#1}{\urlprefix }}%
\providecommand \urlprefix  [0]{URL }%
\providecommand \Eprint [0]{\href }%
\providecommand \doibase [0]{http://dx.doi.org/}%
\providecommand \selectlanguage [0]{\@gobble}%
\providecommand \bibinfo  [0]{\@secondoftwo}%
\providecommand \bibfield  [0]{\@secondoftwo}%
\providecommand \translation [1]{[#1]}%
\providecommand \BibitemOpen [0]{}%
\providecommand \bibitemStop [0]{}%
\providecommand \bibitemNoStop [0]{.\EOS\space}%
\providecommand \EOS [0]{\spacefactor3000\relax}%
\providecommand \BibitemShut  [1]{\csname bibitem#1\endcsname}%
\let\auto@bib@innerbib\@empty
%</preamble>
\bibitem [{\citenamefont {{Zheng}}\ \emph {et~al.}(2011)\citenamefont
  {{Zheng}}, \citenamefont {{Hao}},\ and\ \citenamefont {{Long}}}]{ZHL11}%
  \BibitemOpen
  \bibfield  {author} {\bibinfo {author} {\bibfnamefont {C.}~\bibnamefont
  {{Zheng}}}, \bibinfo {author} {\bibfnamefont {L.}~\bibnamefont {{Hao}}}, \
  and\ \bibinfo {author} {\bibfnamefont {G.~L.}\ \bibnamefont {{Long}}},\
  }\href@noop {} {\bibfield  {journal} {\bibinfo  {journal} {ArXiv e-prints}\ }
  (\bibinfo {year} {2011})},\ \Eprint {http://arxiv.org/abs/1105.6157}
  {arXiv:1105.6157 [quant-ph]} \BibitemShut {NoStop}%
\bibitem [{\citenamefont {Carlini}\ \emph {et~al.}(2006)\citenamefont
  {Carlini}, \citenamefont {Hosoya}, \citenamefont {Koike},\ and\ \citenamefont
  {Okudaira}}]{CHKO06}%
  \BibitemOpen
  \bibfield  {author} {\bibinfo {author} {\bibfnamefont {A.}~\bibnamefont
  {Carlini}}, \bibinfo {author} {\bibfnamefont {A.}~\bibnamefont {Hosoya}},
  \bibinfo {author} {\bibfnamefont {T.}~\bibnamefont {Koike}}, \ and\ \bibinfo
  {author} {\bibfnamefont {Y.}~\bibnamefont {Okudaira}},\ }\href {\doibase
  10.1103/PhysRevLett.96.060503} {\bibfield  {journal} {\bibinfo  {journal}
  {Phys. Rev. Lett.}\ }\textbf {\bibinfo {volume} {96}},\ \bibinfo {pages}
  {060503} (\bibinfo {year} {2006})}\BibitemShut {NoStop}%
\bibitem [{\citenamefont {Bender}\ \emph {et~al.}(2007)\citenamefont {Bender},
  \citenamefont {Brody.}, \citenamefont {Jones},\ and\ \citenamefont
  {Meister}}]{BBJM06}%
  \BibitemOpen
  \bibfield  {author} {\bibinfo {author} {\bibfnamefont {C.~M.}\ \bibnamefont
  {Bender}}, \bibinfo {author} {\bibfnamefont {D.~C.}\ \bibnamefont {Brody.}},
  \bibinfo {author} {\bibfnamefont {H.~F.}\ \bibnamefont {Jones}}, \ and\
  \bibinfo {author} {\bibfnamefont {B.~K.}\ \bibnamefont {Meister}},\ }\href
  {\doibase 10.1103/PhysRevLett.98.040403} {\bibfield  {journal} {\bibinfo
  {journal} {Phys. Rev. Lett.}\ }\textbf {\bibinfo {volume} {98}},\ \bibinfo
  {pages} {040403} (\bibinfo {year} {2007})}\BibitemShut {NoStop}%
\bibitem [{\citenamefont {Mostafazadeh}(2007{\natexlab{a}})}]{M07}%
  \BibitemOpen
  \bibfield  {author} {\bibinfo {author} {\bibfnamefont {A.}~\bibnamefont
  {Mostafazadeh}},\ }\href {\doibase 10.1103/PhysRevLett.99.130502} {\bibfield
  {journal} {\bibinfo  {journal} {Phys. Rev. Lett.}\ }\textbf {\bibinfo
  {volume} {99}},\ \bibinfo {pages} {130502} (\bibinfo {year}
  {2007}{\natexlab{a}})}\BibitemShut {NoStop}%
\bibitem [{\citenamefont {{Masillo}}(2011)}]{Mas11}%
  \BibitemOpen
  \bibfield  {author} {\bibinfo {author} {\bibfnamefont {F.}~\bibnamefont
  {{Masillo}}},\ }\href@noop {} {\bibfield  {journal} {\bibinfo  {journal}
  {ArXiv e-prints}\ } (\bibinfo {year} {2011})},\ \Eprint
  {http://arxiv.org/abs/1105.3332} {arXiv:1105.3332 [quant-ph]} \BibitemShut
  {NoStop}%
\bibitem [{\citenamefont {Mostafazadeh}(2007{\natexlab{b}})}]{M07d}%
  \BibitemOpen
  \bibfield  {author} {\bibinfo {author} {\bibfnamefont {A.}~\bibnamefont
  {Mostafazadeh}},\ }\href {\doibase 10.1016/j.physletb.2007.04.064} {\bibfield
   {journal} {\bibinfo  {journal} {Phys. Lett. B}\ }\textbf {\bibinfo {volume}
  {650}},\ \bibinfo {pages} {208} (\bibinfo {year} {2007}{\natexlab{b}})},\
  \Eprint {http://arxiv.org/abs/0706.1872} {arXiv:0706.1872 [quant-ph]}
  \BibitemShut {NoStop}%
\bibitem [{\citenamefont {Scolarici}\ and\ \citenamefont
  {Solombrino}(2003)}]{SS03}%
  \BibitemOpen
  \bibfield  {author} {\bibinfo {author} {\bibfnamefont {G.}~\bibnamefont
  {Scolarici}}\ and\ \bibinfo {author} {\bibfnamefont {L.}~\bibnamefont
  {Solombrino}},\ }\href {\doibase 10.1063/1.1609031} {\bibfield  {journal}
  {\bibinfo  {journal} {J. Math. Phys.}\ }\textbf {\bibinfo {volume} {44}},\
  \bibinfo {pages} {4450} (\bibinfo {year} {2003})}\BibitemShut {NoStop}%
\bibitem [{\citenamefont {Mostafazadeh}(2002)}]{M02a}%
  \BibitemOpen
  \bibfield  {author} {\bibinfo {author} {\bibfnamefont {A.}~\bibnamefont
  {Mostafazadeh}},\ }\href {\doibase 10.1063/1.1489072} {\bibfield  {journal}
  {\bibinfo  {journal} {J. Math. Phys.}\ }\textbf {\bibinfo {volume} {43}},\
  \bibinfo {pages} {3944} (\bibinfo {year} {2002})}\BibitemShut {NoStop}%
\end{thebibliography}%

\end{document}